# Agentic Observability: Automated Alert Triage for Adobe E-Commerce

Aprameya Bharadwaj[*], Kyle Tu[*]

Adobe
345 Park Ave
San Jose, CA, USA
abharadwaj@adobe.com, ktu@adobe.com

## Abstract

Modern enterprise systems exhibit complex interdependen-cies that make observability and incident response increas-ingly challenging. Manual alert triage, which typically in-volves log inspection, API verification, and cross-referencing operational knowledge bases, remains a major bottleneck in reducing mean recovery time (MTTR). This paper presents an agentic observability framework deployed within Adobe's e-commerce infrastructure that autonomously performs alert triage using a ReAct paradigm. Upon alert detection, the agent dynamically identifies t he a ffected s ervice, retrieves and analyzes correlated logs across distributed systems, and plans context-dependent actions such as handbook consulta-tion, runbook execution, or retrieval-augmented analysis of recently deployed code. Empirical results from production deployment indicate a 90% reduction in mean time to insight compared to manual triage, while maintaining comparable di-agnostic accuracy. Our results show that agentic AI enables an order-of-magnitude reduction in triage latency and a step-change in resolution accuracy, marking a pivotal shift toward autonomous observability in enterprise operations.

## Introduction

Ensuring reliability in large-scale enterprise systems re-quires rapid and accurate diagnosis of production incidents. In modern e-commerce platforms, interdependent microser-vices continuously generate vast streams of operational telemetry that must be interpreted to identify the root cause of failures. Despite advances in observability tooling, inci-dent triage remains largely manual—engineers query logs, inspect dashboards, consult runbooks, and reason across het-erogeneous data sources under intense time pressure. This human-centered workflow contributes substantially to mean time to recovery (MTTR) and constrains situational aware-ness, as critical context is fragmented across tools and data silos. On-call personnel often operate with incomplete in-formation when making recovery decisions. An effective au-tomation framework must therefore accelerate diagnosis and surface context-rich insights, enabling engineers to reason confidently about both cause and impact.

Several LLM-based approaches automate post-failure reasoning. RCACopilot (Chen et al. 2024) and IRCopilot (Lin et al. 2025) perform structured root cause analysis after incidents are registered. RCACopilot maps incidents to pre-defined handlers and generates explanatory narratives, while IRCopilot extends this with a multi-agent design—Planner, Generator, Analyst, and Reflector—for coordinated post-failure diagnosis and remediation. StepFly (Mao et al. 2025) and FLASH (Zhang et al. 2024) focus on executing prede-fined troubleshooting workflows, transforming unstructured guides into DAGs and leveraging historical failures for re-curring incidents. While these systems improve post-alert reasoning or workflow execution, they operate reactively and presuppose fault categorization, lacking integration with live telemetry or proactive hypothesis generation. In con-trast, our framework performs preemptive triage at the mo-ment of alert, continuously correlating alert context with live operational telemetry to generate actionable hypotheses and recommend interventions. This shifts the locus of automa-tion from postmortem analysis to live incident reasoning, narrowing the gap between detection and insight.

TRIANGLE (Yu et al. 2025) and related multi-agent sys-tems address semantic heterogeneity and collaboration in in-cident data, improving interpretability for post-failure triage. However, they do not provide real-time reasoning or inte-grate tightly with observability platforms. Our system in-troduces task-specialized functional agents—including Re-flection, Tools, and telemetry retrieval agents—that unify reasoning, evidence collection, and remediation in a single workflow.

Beyond system-specific architectures, conceptual frame-works such as Collaborative LLM Agents for High-Stakes Alert Triage (Wei et al. 2025) and A Practical Framework for Developing Agentic AIOps Systems (Zota, Bărbulescu, and Constantinescu 2025) emphasize role specialization, planning, and tool grounding, but remain largely theoret-ical and unvalidated in production environments. Holis-tic AI-driven Network Incident Management (Hamadanian et al. 2023) further highlights that no one-size-fits-all solu-tion exists: errors, telemetry structures, and recovery work-flows are domain-specific and require context-aware adapta-tion. Our framework addresses these limitations by integrat-ing domain-specific observability tools, deployment meta-data, and telemetry from Adobe's e-commerce ecosystem, enabling real-time, context-grounded reasoning tailored to service-specific failure patterns.

In summary, while prior systems focus on post-failure analysis, structured workflow execution, or conceptual design principles, our work advances the field through a production-grade, agentic observability framework that performs proactive triage, unifies heterogeneous enterprise data, and demonstrates measurable operational impact—achieving over 90% reduction in Mean Time to Insight (MTTI) while maintaining human-level diagnostic accuracy.

This work makes the following key contributions:

- **Proactive, real-time alert triage:** Unlike existing RCA and troubleshooting frameworks that act post-failure, our system performs diagnostic reasoning as alerts are raised, enabling earlier mitigation and eliminating post-failure latency.
- **Cross-domain, context-aware reasoning:** Prior approaches rely on structured incident data or predefined workflows. Our framework integrates heterogeneous enterprise sources—Splunk logs, distributed traces, deployment metadata, and documentation—enabling reasoning over unstructured telemetry across services.
- **Production-validated impact:** While most prior works remain conceptual or prototype-level, our agentic framework is deployed in Adobe's production e-commerce observability stack, achieving a **90% reduction in Mean Time to Insight (MTTI)** while maintaining expert-level diagnostic accuracy.

## Proposed Method

### System Design

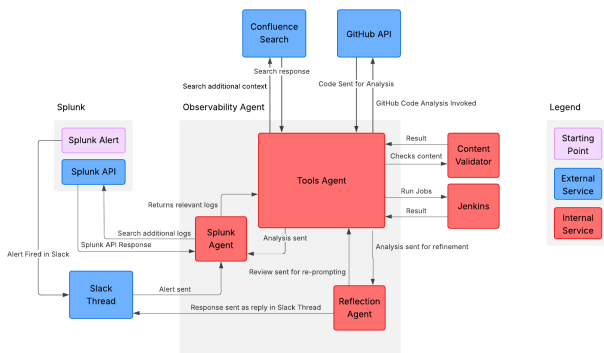

Figure 1: System Design of Observability Agent

Incident triage in enterprise-scale systems involves iterative and interdependent tasks such as: log retrieval, causal reasoning, and validation. While the triage process remains sequential in its logical flow, the multi-agent structure enables interleaved parallelism once the initial evidence is established. For example, after the Tools Agent identifies affected dependencies, the Splunk Agent can asynchronously fetch logs from downstream services while the Tools Agent executes diagnostic reasoning or tool invocations. This overlap reduces total triage time and prevents idle waiting between stages.

The proposed observability framework orchestrates alert triage through three specialized GPT-4o agents—**Splunk Agent**, **Tools Agent**, and **Reflection Agent**—coordinated via *LangGraph*. This multi-agent design allows the system to emulate the decision-making workflow of experienced on-call engineers while maintaining scalability and consistency across multiple services.

**Alert Reception and Log Retrieval** When an alert is triggered from the monitoring system, it is delivered to the Splunk Agent, which extracts relevant metadata including the affected service, alert type, and identifiers such as `sessionId` or `requestId`. Using this information, the Splunk Agent queries the Splunk API to collect logs relevant to the alert, filters for error-level events, and traces identifiers through downstream services to correlate distributed events. The output is a structured dataset of anomalies and candidate causal events, forming the foundation for subsequent reasoning.

**Planning and Reasoning** The Tools Agent acts as the planner and reasoning engine. It interprets the alert metadata and structured log summaries from the Splunk Agent, identifying information gaps and defining sub-goals such as verifying API responses, validating recent deployments, or triggering auxiliary checks. Using a reasoning-trace schema, the Tools Agent formulates a stepwise action plan that specifies which tools or knowledge sources to invoke, in what order, and under what conditions. Retrieval-augmented generation (RAG) is used to integrate information from runbooks, internal wikis, and the latest code deployment metadata, allowing the agent to generate grounded explanations and actionable recommendations.

**Action Execution** Once the plan is defined, the Tools Agent can autonomously execute remediation tasks when applicable. Human oversight is maintained for high-risk operations to ensure operational safety and compliance. Combining reasoning with actionable tool invocation, the system reduces manual effort and accelerates the triage process.

**Reflection and Feedback Loop** The Reflection Agent performs a meta-evaluation of the synthesized findings to ensure that the outputs satisfy three key criteria: **completeness** (all relevant error paths and affected services are covered), **causality** (the inferred root cause is logically supported by log evidence), and **actionability** (recommended resolutions are operationally feasible). To prevent indefinite reasoning loops—a potential issue when confidence thresholds are not met—the system imposes a maximum of five reflection cycles. If the Reflection Agent still detects uncertainty beyond this limit, it terminates the loop and returns the most confident hypothesis generated by the Tools Agent, along with an explicit uncertainty tag. This safeguard ensures responsiveness and bounded compute time while maintaining interpretability in low-confidence scenarios.

**Dynamic Knowledge Retrieval and Action Integration** The Tools Agent also dynamically decides which knowledge sources to query based on its inferred root-cause hypotheses. For example, if the agent's analysis suggests a code re-gression as the likely cause of an observed error, it performs retrieval-augmented generation (RAG) over recent code deployment metadata to identify the most probable commits or changes responsible for the issue. Similarly, contextual information from runbooks or internal documentation is retrieved only when relevant to the agent's reasoning path.

Once relevant evidence is retrieved and synthesized, the agent can propose or autonomously execute actionable remediation steps—such as restarting jobs, clearing caches, or running validation scripts—with human confirmation where necessary. This dynamic coupling of reasoning, targeted knowledge retrieval, and conditional action ensures that each alert is triaged efficiently, focusing only on relevant information and minimizing unnecessary queries or operations.

# Evaluation

This section presents an empirical evaluation of the proposed agentic observability framework in the context of Adobe's e-commerce infrastructure. The evaluation focuses on triage efficiency, information completeness, and diagnostic accuracy, comparing the system's performance against existing manual workflows used by on-call engineers and support analysts.

## Experimental Setting

The agent was deployed across multiple production services within the Adobe e-commerce ecosystem, each instrumented with standard alerting and logging pipelines. Prior to deployment, incident triage was conducted manually by engineers or by a dedicated support team using Splunk dashboards, internal handbooks, and runbooks.

During a twelve-week observation period, we collected data from 250 alert events across diverse service categories, including checkout, subscription management, and catalog ingestion. For each event, we recorded timestamps from alert trigger to actionable diagnosis, the set of data sources consulted, and the final root cause determined through post-incident review.

## Baselines

We define two baselines corresponding to existing operational practices:

**Manual Engineer Triage:** On-call engineers who are domain experts perform log searches and knowledge lookups manually using Splunk and internal documentation. Average triage time: 18 minutes.

**Manual Support Triage:** Support analysts form a centralized team responsible solely for monitoring alerts across all services. Since they lack deep service-specific context, they perform only initial diagnosis and escalate to engineers when necessary. Their average triage time is about 33 minutes, reflecting broader alert coverage but lower diagnostic precision.

These baselines represent the status quo of incident handling, against which the proposed agent's performance is compared.

| Metric | Manual Engineer | Manual Support | Agentic Support |
|---|---|---|---|
| MTTI (minutes) | 18.4 | 33.2 | 2.3 |
| ELA | 82.4% | 55.6% | 88.4% |
| EER | N/A | 15% | 65% |
| AR | 65.2% | 48.4% | 90.4% |

Table 1: Metrics for all alerts over a 12-week span

## Evaluation Metrics

To quantify improvements, we employ five metrics that capture time, quality, and completeness dimensions of the triage process:

**Mean Time to Insight (MTTI):** The elapsed time from alert trigger to generation of an actionable diagnostic summary.

**Error Localization Accuracy (ELA):** The proportion of alerts where the system's predicted fault component matches the verified root cause identified post-resolution.

**Engineer Effort Reduction (EER):** The ratio of triage steps automated by the agent relative to the total steps required manually.

**Alert Responsiveness (AR):** The percentage of alerts for which an initial diagnostic report was produced within five minutes of trigger.

The results, summarized in Table 1, represent averages across all alerts observed over a 12-week period. For the manual engineer and support comparisons, we analyzed the same number of similar alerts that occurred prior to the deployment of the agent. The agentic system achieves a tenfold reduction in Mean Time to Insight (MTTI), delivering actionable diagnostic summaries within approximately two minutes, compared to 18–30 minutes for manual triage. Its error localization accuracy of 88.4% is comparable to expert engineers, though lower for general on-call engineers due to variability in individual expertise, while operating with substantially higher speed and consistency. By automating 65% of manual triage steps (Engineer Effort Reduction), the system significantly reduces cognitive load, allowing engineers to focus on corrective actions. Alert Responsiveness reaches approximately 90%, with minor deviations primarily caused by temporary Splunk API throttling.

## Case Study: Automated Diagnosis of Content Validation Errors

To evaluate the agent's real-world effectiveness, we examine a production incident corresponding to the alert "Splunk Alert: Content Validation Error – WARN." This alert is triggered when the Adobe Experience Manager (AEM) service returns content containing invalid or malformed string formats. The alert is configured to fire every minute until the malformed content is corrected, resulting in a rapid accumulation of alerts within a short time frame. Unlike typical alerts where context can be reused, each instance of this alert must be independently triaged because multiple fragments across different resources may contain distinct formatting errors. Consequently, prior diagnostic outputs are not reusable across alerts, making this scenario a rigorous test of the agent's ability to perform rapid, context-specific reasoning at scale. Such anomalies are critical because, upon detection, the checkout platform immediately falls back to static , cached content served from the CDN and does not retrieve the latest versions.

### Incident Context

In traditional workflows, the on-call engineer must manually:

- Inspect Splunk logs to locate the affected content fragment (5-8 minutes).
- Identify the specific variant and locale involved (1-2 minutes).
- Execute the content validation script to determine the precise line number and nature of the formatting error (2-3 minutes).
- Apply corrections in AEM and re-publish the validated content (1-2 minutes).

This process typically requires 10-15 minutes, depending on system load and log volume, and demands contextual switching across multiple dashboards and internal tools.

### Agentic Resolution Flow

Upon receiving the alert webhook, the Splunk Agent initiated contextual log retrieval for the associated session identifiers (30-45 seconds). The retrieved logs were parsed to isolate the malformed fragment, after which the Tool Agent inferred the most probable variant and locale based on request metadata and prior deployment history (20-30 seconds). The Tools Agent invoked the internal content validation script, obtaining both the line number and error description for the faulty entry (20-30 seconds). The agent synthesized these results into a structured diagnostic summary posted to Slack after the Reflection agent validates the response of the Tools agent. The response specifies the Fragment name, Variant and locale, Error type and line number, and Recommended corrective action

This alert was triggered 72 times over a 12-week period. For comparison, Table 2 analyzes the 72 most recent occurrences prior to the deployment of our system to represent manual on-call performance. The agent reduced the engineer's effort, now the only manual step is modifying and re-publishing the affected content, significantly lowering both cognitive and operational load. The system generated initial diagnostic summaries for 91.6% of alerts—compared to only 61.1% coverage under manual triage despite the high frequency and repetitive nature of these alerts. By automating three of the four manual triage steps, the workflow achieves an Engineer Effort Reduction (EER) of 75%.

## Discussion and Future Work

These findings indicate that the proposed agentic framework effectively replicates and accelerates the cognitive workflow of human operators. The improvements in Mean Time to Triage (MTTI) and responsiveness directly translate to lower Mean Time to Resolution (MTTR) and enhanced service reliability. Importantly, the observed parity in diagnostic accuracy suggests that LLM-driven reasoning, when grounded in enterprise data sources, can achieve human-level interpretability without sacrificing precision.

| Metric | Manual On-Call Engineer | Agent Assisted Engineer |
|---|---|---|
| MTTI (minutes) | 13.3 | 1.8 |
| EER | N/A | 75% |
| AR | 61.1% | 91.6% |

Table 2: Comparison metrics of case study

However, the evaluation also highlights several current limitations. The system's reasoning quality remains dependent on the fidelity of log data and the completeness of onboarded services. In certain high-alert periods, Splunk API rate limits can throttle log retrieval, causing the agent to experience temporary response delays and, in rare cases, no log retrieval. While these slowdowns do not affect reasoning correctness, they reduce responsiveness under peak-load conditions. Additionally, although automated tool invocation yields substantial efficiency gains, it necessitates strict access control and human oversight for high-risk remediation tasks. Supporting new alerts also requires manual onboarding of runbooks, wikis, and tools—a process that remains human-dependent—though the framework provides a structured foundation to streamline this process.

Future work aims to incorporate uncertainty estimation and feedback-driven learning to further improve diagnostic robustness, while exploring asynchronous caching or incremental log indexing to mitigate API throttling effects. More broadly, this work illustrates how agentic AI systems can operationalize reasoning, retrieval, and action in enterprise observability—transforming reactive monitoring pipelines into proactive, self-correcting workflows.